\documentstyle[preprint,eqsecnum,aps,prb]{revtex}
\begin{document}
\title{Structural and magnetic properties of a series of low doped Zn$_{1-x}$Co$_x$%
O thin films deposited from Zn and Co metal targets on (0001) Al$_2$O$_3$
substrates.}
\author{A.\ Fouchet, W. Prellier\thanks{%
prellier@ismra.fr}, P.\ Padhan, Ch.\ Simon and B. Mercey}
\address{Laboratoire CRISMAT, CNRS\ UMR 6508, 6 Bd du Marechal Juin,\\
F-14050 Caen Cedex, FRANCE.}
\author{V.N.\ Kulkarni\thanks{%
on leave from Indian Institute of Technology, Kanpur, India} and T.\
Venkatesan}
\address{Center for Superconductivity Research, University of Maryland,\\
College Park, MD\ 20742-4111, USA}
\date{\today}
\maketitle

\begin{abstract}
We report on the synthesis of low doping Zn$_{1-x}$Co$_x$O ($0<x<0.1$) thin
films on (0001)-Al$_2$O$_3$ substrates. The films were prepared in an
oxidizing atmosphere, using the pulsed laser deposition technique starting
from Zn and Co metallic targets. We first studied the influence of the
strains of ZnO and their stuctural properties. Second, we have investigated
the structural and the magnetic properties of the Zn$_{1-x}$Co$_x$O films.
We show that at low doping, the lattice parameters and the magnetization of
the Zn$_{1-x}$Co$_x$O films depend strongly on the Co concentration.
\end{abstract}

\newpage

\section{Introduction}

Diluted Magnetic Semiconductors (DMS) of III-V or II-VI types have been
obtained by doping semiconductors with magnetic impurities (Mn for example)%
\cite{1,2}. These materials are very interesting due to their potential
applications for spintronics\cite{3}. However, the low Curie temperature ($%
T_C$) has limited their interest\cite{4} (for example Ga$_{1-x}$Mn$_x$As,
with $x=5.3\%$ has a $T_C=110K$\cite{LTC}). Based on the theoretical works
of Dietl {\it et al.}\cite{5}, several groups \cite{5a} have studied the
growth of Co-doped ZnO films\cite{6,7,8,9} which is a good candidate having
a high $T_C$ \cite{5}. Using pulsed laser depositions (PLD), Ueda {\it et al.%
} reported ferromagnetism (FM) above room temperature\cite{6}, while Jin 
{\it et al.} found no indication of FM by utilizing laser molecular beam
epitaxy\cite{7}. This controversy between research teams may result from the
growth method used and/or from the growth conditions (oxygen pressure,
deposition temperature, etc...). In the particular case of the PLD
technique, it may also arise from the targets preparation and this parameter
has never been considered up to now. One of the reason is that the control
of the dopant incorporation would be quite difficult to obtain using a
pre-doped ceramic oxide target\cite{10}. This is a crucial point since the
properties of the DMS are very sensitive to the percentage of dopant\cite{11}%
. The homogeneity of dopant incorporation as well as the precise control of
the growth might be responsible for the changes in the physical properties
of the films obtained by the different groups. Thus, we have recently
developed an accurate method to grow the Zn$_{1-x}$Co$_x$O films with a
precise doping by using an alternate deposition starting from Zn and Co
targets. Using this procedure\cite{12}, we have been able to observe
ferromagnetism at room temperature in one sample having the composition Zn$%
_{0.95}$Co$_{0.05}$O confirming previous results\cite{6} but in contrast to
others\cite{7}. In this paper, we have grown a series of Zn$_{1-x}$Co$_x$O
films with very low doping of Co. We have studied the structural and
magnetic properties of the films and our results are reported in this
communication.

\section{Experimental}

The Zn$_{1-x}$Co$_x$O films were grown using the pulsed laser deposition
technique. Metal Zinc ($99.995\%$) and Cobalt ($99.995\%$) targets were used
as purchased (NEYCO, France) without further preparations. The films are
deposited using a KrF laser ( $\lambda =248mm$)\cite{12} on (0001) Al$_2$O$%
_3 $ substrates (in this paper, we used the three-index notation instead of
the four-index one\cite{Hexa}). The substrates were kept at a constant
temperature in the range 500${{}^{\circ }}$C-750${{}^{\circ }}$C during the
deposition which was carried out a pressure around $0.1Torr$ of pure oxygen.
After deposition, the samples were slowly cooled to room temperature at a
pressure of $225Torr$ of O$_2$. The deposition rate is $3Hz$ and the energy
density is close to $2J/cm^2$. The composition of the film was checked by
Energy Dispersive Scattering and Rutherford Backscattering Spectometry (RBS).

The structural study was done by X-Ray diffraction (XRD) using a seifert XRD
3000P for the $\Theta -2\Theta $ scans and the $\omega -$scan to evaluate
the full-width-at-maximum (FWHM). An X'Pert Phillips was utilized for the 
{\it in-plane} measurements obtained from the ($101$) reflection. Ion
Channeling technique (with 2 MeV He$^{+}$ ions) was used to study the
epitaxial nature of the films.

Magnetization ($M$) was recorded as a function of the temperature ($T$) and
the magnetic field ($H$) in a SQUID\ magnetometer.

\section{Results}

ZnO thin film was deposited using the optimal conditions found previously%
\cite{12}. The films are highly crystallized as seen from the sharp
diffraction peaks (Fig.1) and the FWHM of the rocking curve close to 0.25${%
{}^{\circ }}$ (Fig.4). The two diffractions peaks observed around $34.48{%
{}^{\circ }}$ and $72.66{{}^{\circ }}$ are characteristic of the hexagonal
ZnO wurtzite, the $c$-axis being perpendicular to the substrate plane. The
out-of-plane lattice parameter is calculated to be $0.52nm$ which
corresponds to the theoretical bulk one\cite{13}. The epitaxial
relationships between ZnO films and Al$_2$O$_3$ substrates are determined
using asymmetrical XRD. The inset of Fig.1 displays the $\Phi -$scan both of
the ZnO films obtained from the ($101$) planes and the Al$_2$O$_3$ substrate
obtained from the ($104$) planes. The peaks belonging to the film are
separated by 60${{}^{\circ }}$, indicating a six-fold symmetry in agreement
with the hexagonal structure of ZnO whereas the peaks of the sapphire
substrate are separated of 120${{}^{\circ }}$ indicating a three-fold
symmetry in agreement with the rhombohedral symmetry of Al$_2$O$_3$. Note
that 30${{}^{\circ }}$ spacing between the diffraction peaks of the film and
those of the substrate indicate a rotation of 30${{}^{\circ }}$ between the
in-plane axes. In order to obtain additional information, on the structural
properties of the ZnO films, we have determined the strains of the optimized
film. This technique used the distance between atomic plane of a crystalline
specimen as an internal strain gage\cite{15}. Using this model, we minimized
the strains of the ZnO film by changing the growth conditions \cite{12}. At
a temperature of 600${{}^{\circ }}C$ under $0.1Torr$ of O$_2$, we found\cite
{15B} that the value of residual stress ($\sigma $) along the in{\it -plane}
direction ($\sigma _\Phi =150MPa$) is about the same value along the {\it %
out--of-plane} direction ($\sigma _{\bot }=150MPa$). Such values are in
agreement with previous report \cite{Perriere}. The study of the ($10l$)
planes indicates a concentration gradient along the {\it out-of-plane}
direction.\ In other words, the strains is larger near the interface and
decreases along the {\it out-of-plane} direction as the thickness increases.
Indeed, when the thickness of the film, the substrate-induced strain becomes
lower and the lattice parameters of films become closer to the value of the
bulk. Moreover, the positive slope\cite{15B} indicates an extensive stress 
{\it in-the-plane} of the substrate which is in agreement\cite{Wil} with the
decrease of the {\it out-of-plane} lattice parameter as compared to the bulk
value\cite{13}.

These conditions have been used to grow the Zn$_{1-x}$Co$_x$O films with
various doping x.\ An example of the ion backscattering channeling data for
Zn$_{1-x}$Co$_x$O (x=4\%) is given in Fig.2. The observed minimum yield $%
\chi _{\min }$ of $\approx $5\% for channeling reflects good quality of the
film, thereby indicating optimum growth of Zn$_{1-x}$Co$_x$O films. Note
also that a surface peak of Co is seen which reveals that the top layer
(about 400$\AA $) has large amount of Co ($11\%$). The channeling spectrum
also shows the Co signal. This indicates that the Co in this layer has not
gone substitutional. Fig.3 shows the $\Theta -2\Theta $ scans recorded
around the $002$ reflection in the the range $34-35{{}^{\circ }}$. As the Co
content is increasing there is a shift toward the high angle, indicating a
decrease of the {\it out-of-plane} lattice parameter up to $1.6\%$.\ Above
this value, the position of the diffraction peak is moving toward the low
values of 2$\Theta $, leading to an increase of the lattice parameter up to $%
9\%$.\ Then, the {\it out-of-plane} lattice parameter becomes constant\cite
{12} (not show in this graph). To explain this results, we consider the ZnO
bulk wurtzite structure, where zinc (Zn$^{2+}$) is in tetrahedral site with
a ionic radius of $0.6\AA $\cite{Shannon}. The substitution of the Zn$^{2+}$
with Co$^{2+}$ (ionic radius: $0.58\AA $\cite{Shannon}) leads to a
compression of the bulk structure and a decrease of the lattice parameters.\
On the contrary, a higher coordination number of Co$^{2+}$ leads to an ionic
radius of Co$^{2+}$ larger than the ionic radius of Zn$^{2+}$. In this
configuration, Co$^{2+}$ would be in an interstitial site inducing an
expansion of the structure (and an increase of the lattice parameters) and
thus, an increase of defects. In the thin film, due to the substrate-induced
strains, this will lead to a decrease of the {\it out-of-plane} lattice
parameter (up to $1.6\%$ of Co) and then to an increase of this parameter ($%
1.6\%<x<9\%$).\ This explanation (substitution and then interstitial Co) is
also confirmed by looking at the evolution of the FWHM of the rocking curve
as a function of the Co content (see Fig.4) where the same tendancy is
observed. For very low doping of Co ($0<x<1.6\%$), the FWHM is slightly
decreasing up to $1.66\%$ of Co. Above this value ($x>1.6\%$), the FWHM
increases when the Co content is also increasing indicating that for these
values of doping, the films are less oriented and/or contained more defects.

We investigated the magnetic properties of these thin film samples. Fig.5
shows the $M(T)$ recorded for various Zn$_{1-x}$Co$_x$O film. The films at
low doping of Co ($1.6\%$ and $6.6\%$) clearly evidence a ferromagnetic
state with a Curie temperature respectively around $150K$\ and $300K$. For
the other concentration of Co, the situation is more confusing.\ We have
performed $M(H)$ at different temperatures but we did not observed any
hysteresis. We also measure the magnetization in very low doping ($x<1.6\%$)
but the signal of the substrate is too high.

We discuss now the possibility of the Co clusters in our films.\ The
transition from the ferromagnetic state to the paramagnetic state is clearly
seen, suggesting that the metallic Co clusters (the $T_C$\ of the metal Co
clusters is above 1000K) are not responsible for the effect observed at 300K 
\cite{8,9}.\ Moreover, the saturation moment (0.7 $\mu _B$/mole Co) is very
weak compared to $1.7\mu _B$ of metallic Co$^{[0]}$, suggesting that the Co
state should be close to Co$^{2+}$. We believe that this is due to the
technique used in the study where not only the conditions of the deposition
minimize the strains but also the alternate deposition from the two targets
favors the homogeneity of the doped films. Moreover, it has been seen that
the low temperature (600${{}^{\circ }}C$ is our case) of deposition leads to
homogenous films\cite{8}.

\section{Conclusion}

In conclusion, we have developed an alternative method for the growth of
pulsed laser deposited oxide thin films. This method permits an accurate
control of the dopant in the matrix. Firstly using this procedure, we have
deposited high quality Zn$_{1-x}$Co$_x$O thin films on Al$_2$O$_3$ (0001)
substrates with a low doping of Co.\ Secondly, we have also studied the
evolution of the structure and the magnetic properties of the films for
various Co concentration. We suggest that the incorporation of Co inside the
wurtzite structure leads to an increase of interstitial defects while the
epitaxial defects decrease. Finally, the growth of these ferromagnetic films
opens the route for the fabrication of spin-based electronics since this
original method can be used to grow various oxide thin films.

We acknowledge the Centre Franco-Indien pour la Promotion de la Recherche
Avancee/Indo-French Centre for the Promotion of Advance Research
(CEFIPRA/IFCPAR) for financial support (Project N${{}^{\circ }}$2808-1).
A.F. also thanks the CNRS\ and the ''Conseil Regional de Basse Normandie''
for his BDI fellowship.

Figures Captions:

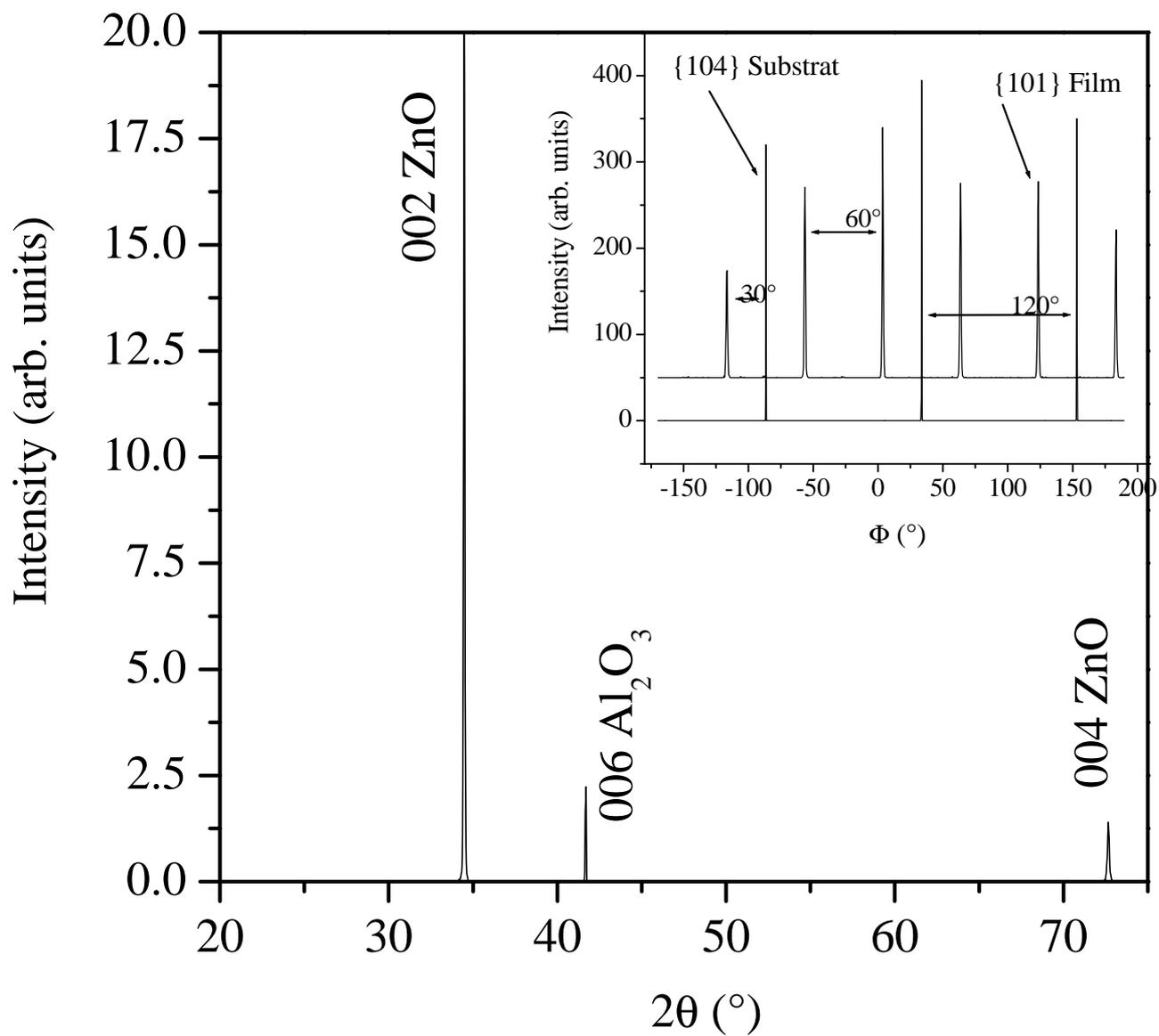
Figure 1: Room temperature $\Theta -2\Theta $ XRD pattern of typical ZnO
film. The inset depicts the $\Phi $-scan of the film and the substrate
respectively recorded around the \{101\} and \{104\} family peak.

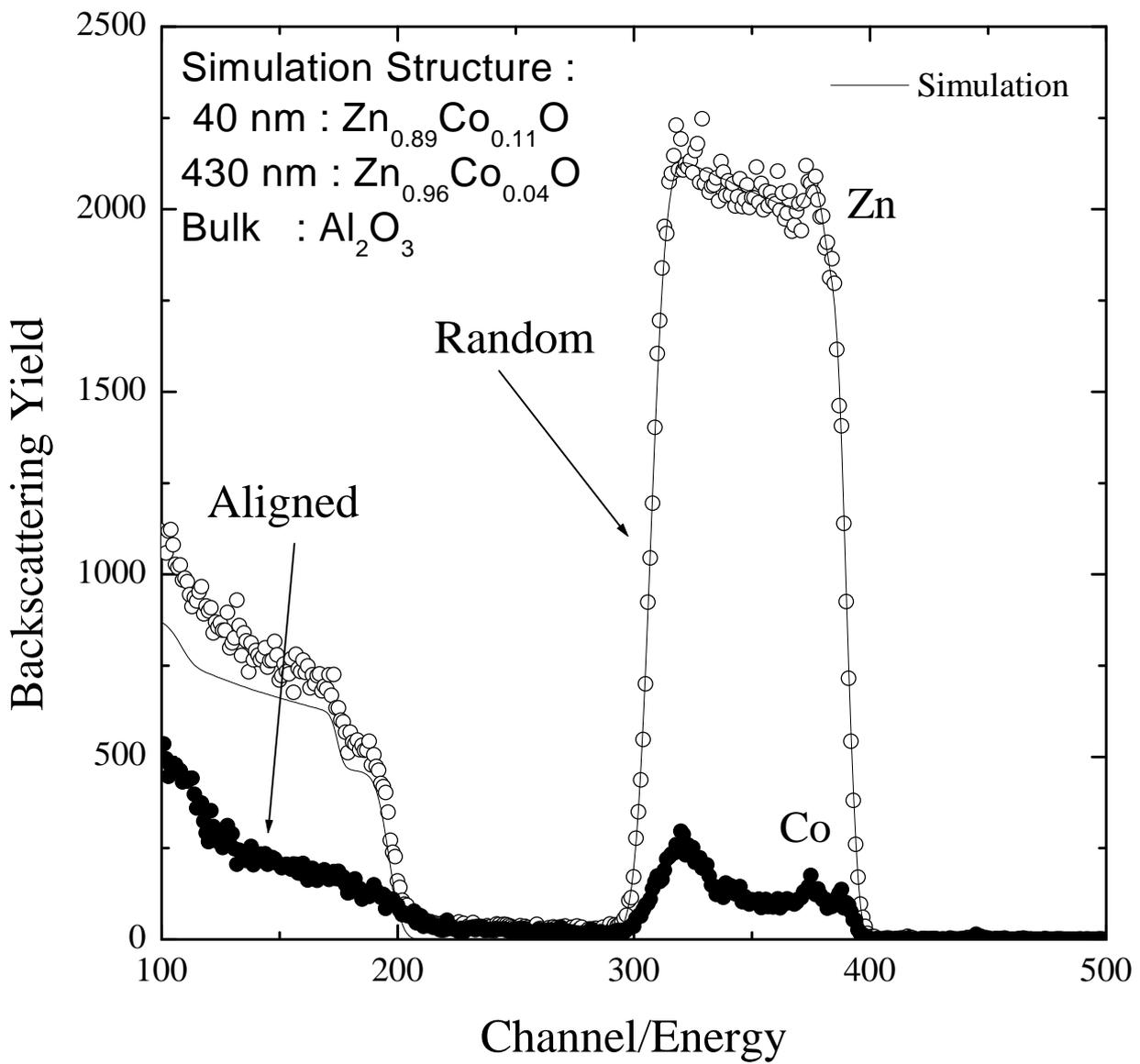
Figure 2: The 2 Mev\ He$^{+}$ Rutherford Backscattering random and channeled
spectra of Zn$_{1-x}$Co$_x$O film (x=4\%).

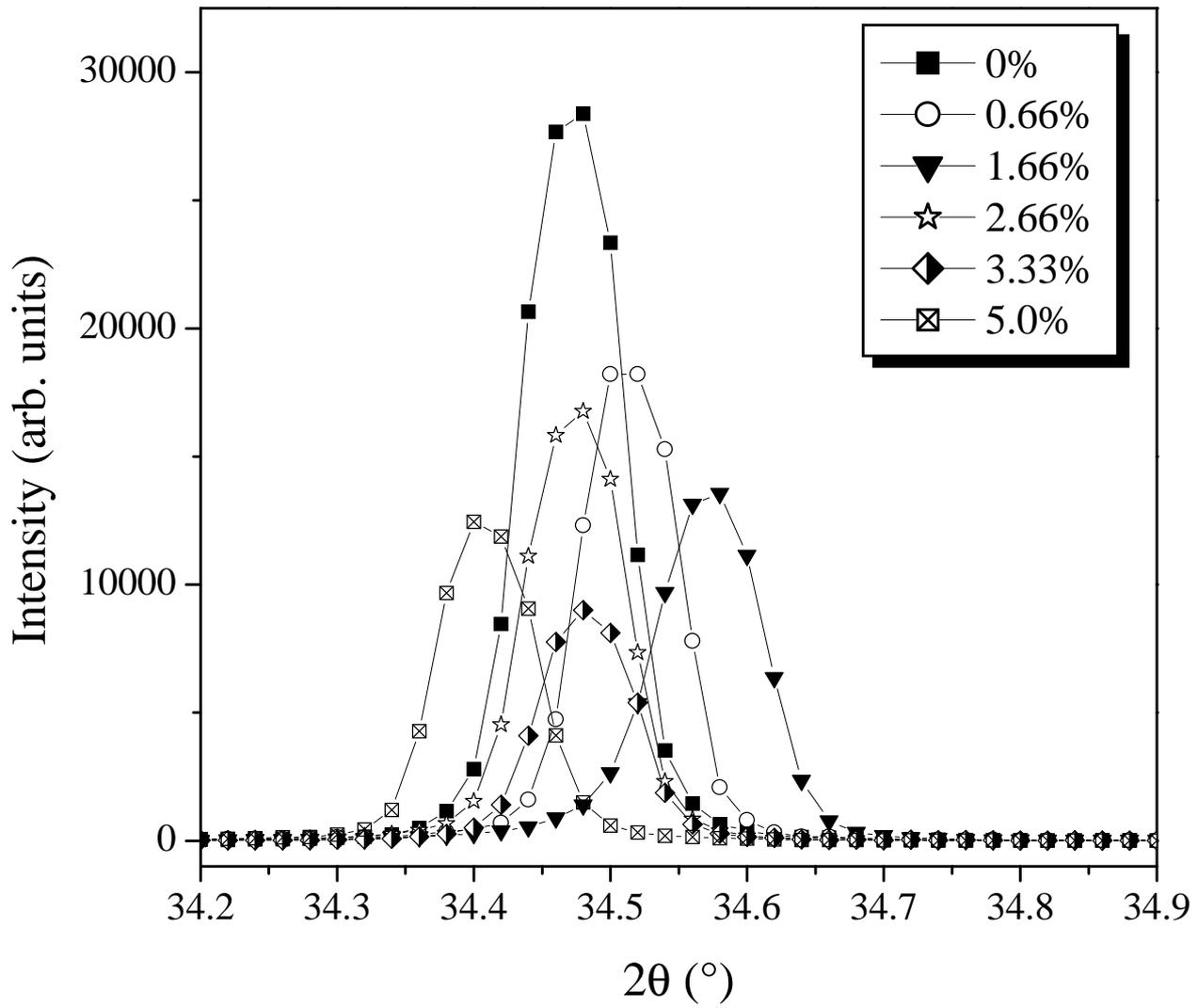
Figure 3: Room temperature $\Theta -2\Theta $ XRD pattern of a series of Zn$%
_{1-x}$Co$_x$O film. The percentage of Co is indicating.

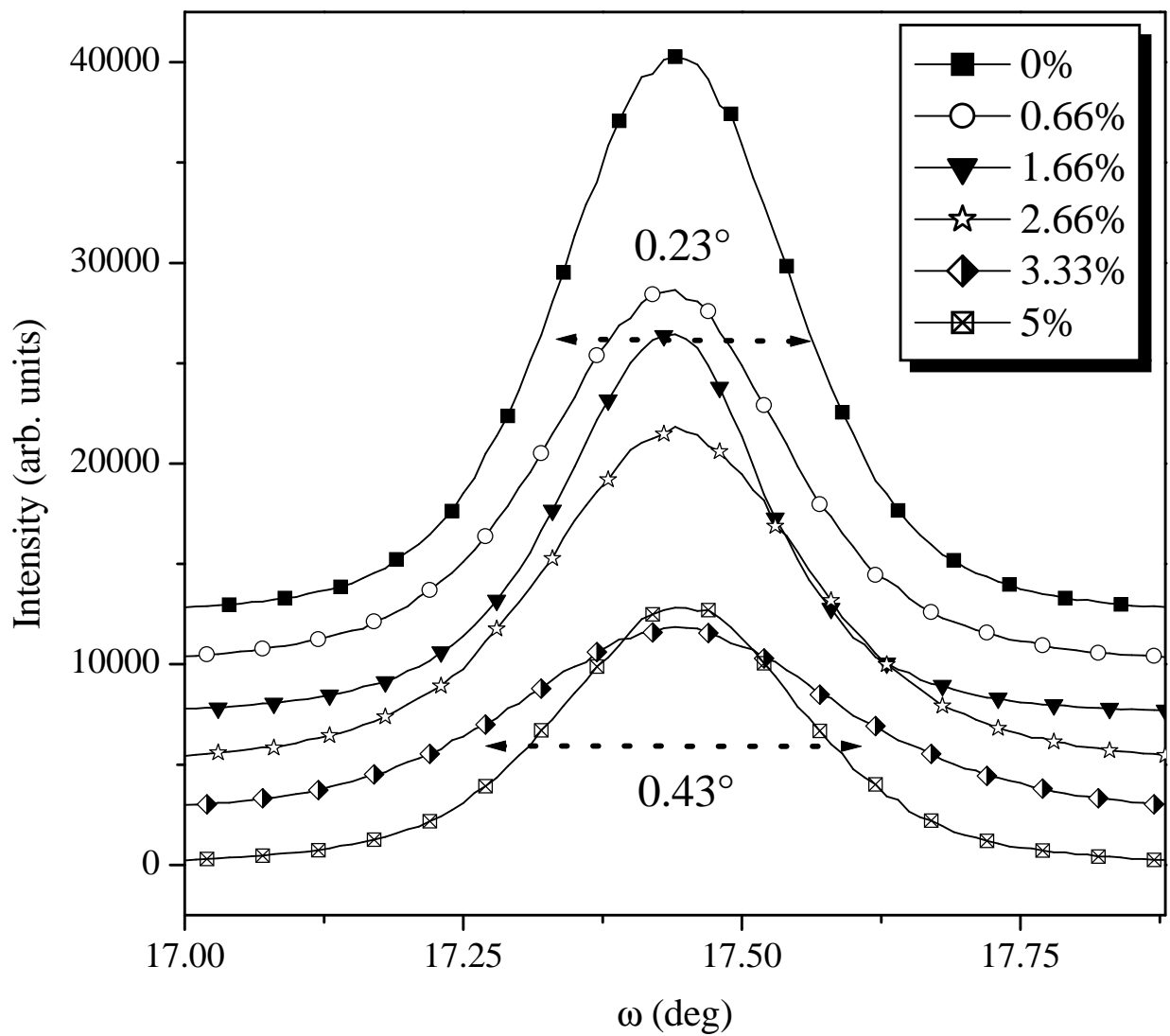
Figure 4: Rocking curve ($\omega $-scan) recorded around the $002$
reflection for a series of Zn$_{1-x}$Co$_x$O film.The percentage of Co is
indicating. The FWHM is increasing from 0.23${{}^\circ}\;$for pure ZnO to
0.43${{}^\circ}$ for Zn$_{1-x}$Co$_x$O with x=5\%.

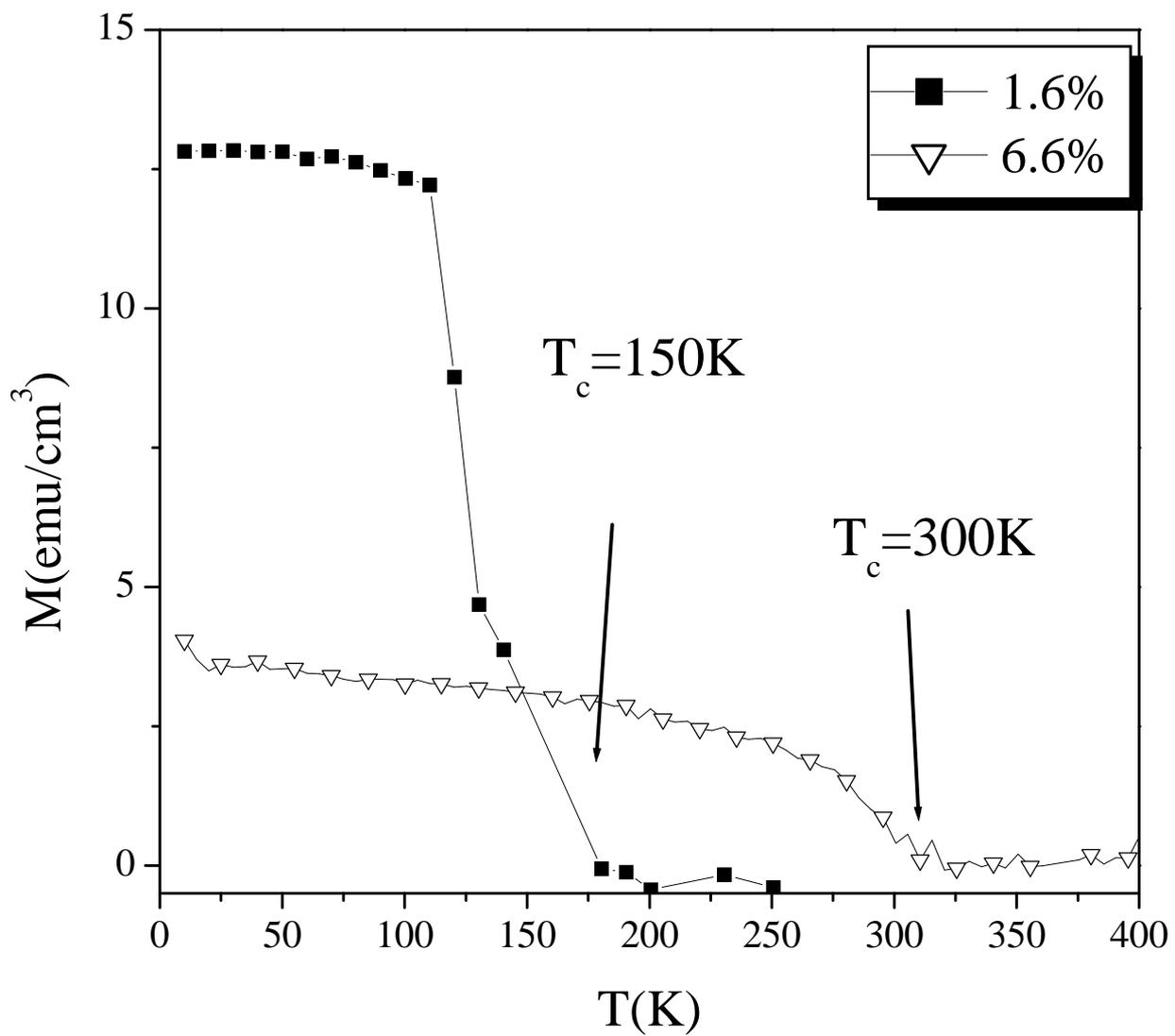
Figure 5: $M(T)$ of a series of Zn$_{1-x}$Co$_x$O film measured sith a field
of $2000Oe$. The percentage of Co is indicating.

\end{document}